\documentstyle[12pt]{article}
\begin{document}
\begin{center}
{\bf Bacterial evolution and the Bak-Sneppen model}
\vskip 1cm
Indrani Bose and Indranath Chaudhuri$^\ddag$
\\Department of Physics, \\Bose Institute,
\\93/1, Acharya Prafulla Chandra Road,
\\Calcutta-700 009, India.
\\$\ddag$ Department of Physics, \\St. Xavier's College,
\\30, Park Street, Calcutta-700016, India.
\end{center}

\begin {abstract}
Recently, Lenski et al \cite{Elena,Lenski,Travisano} have carried out several experiments on bacterial evolution.
Their findings support the theory of punctuated equilibrium in biological
evolution. They have further quantified the relative contributions of adaptation,
chance and history to bacterial evolution. In this Brief Report, we show that a modified
M-trait Bak-Sneppen model can explain many of the experimental results in a
qualitative manner.
\\{bf PACS : 87.23.Kg, 05.65.+b,87.17.Ee}
\end{abstract}

Recently, a set of experiments has been carried out on bacterial evolution
which has given rise to a new sub-discipline in the area of evolutionary
biology, namely, experimental evolution \cite{Elena,Lenski,Travisano,RE}. This implies the study, in
the laboratory, of the fundamental processes of evolutionary change, namely,
spontaneous mutation and adaptation by natural selection. Experiments on
evolutionary dynamics require passage through thousands of generations which
is an impossibility for most living species. Microorganisms like bacteria, yeast
or viruses, however, have very short generation times (bacteria like E.coli have
about seven generations everyday in sugar solution). The short generation times
make it possible to observe population dynamics over thousands of generations 
and thereby address a wide range of evolutionary questions. Darwin laid the
foundation of evolutionary biology by setting forth the principle of adaptation
by species through natural selection \cite{Darwin}. It is now known 
that spontaneous mutation plays an important role in generating differences
among individual organisms. A mutation involves a change in the base sequence 
of a DNA and can occur with a certain probability during cell division.  Mutations are random
events and may be harmful, neutral or beneficial as regards their effect on
an organism. According to the modern version of Darwin's theory, random mutations
give rise to heritable differences among organisms whereas natural selection
tends to increase the number of fitter variants. The processes of reproduction,
mutation and natural selection are responsible for evolution of species or that
of a single population.

In this Letter, we  consider a set of experiments carried out by Lenski 
and co-workers \cite{Elena,Lenski,Travisano} on the bacteria E.coli. As pointed out by them
\cite{T}, a large population size of E.coli ensures that a large number of mutations
occur in every generation so that the origin and the fate of genetic variation can 
be well studied. It is possible to store bacteria in suspended animation at low
temperatures. One can then measure the relative fitness (RF) of descendant and
ancestral populations by placing them in direct competition. The RF is expressed 
as the ratio of the realized growth rates of the two populations. Finally, the
populations are easy to handle and propagate so that intensive replication of 
experiments is possible allowing subtle effects to be measured. In the first
experiment \cite{Elena,Lenski}, Lenski and co-workers studied an evolving bacterial population for
approximately 10,000 generations. The bacterial population was allowed to expand
to $5\times10^{8}$ cells in low sugar solution. At the end of the growth period (one day),
$1/100$ th of the population was siphoned into a fresh flask of food to allow
the population to evolve. Since there was a 100-fold expansion of the bacterial
population in a day, $6.6 (=log_{2}{100})$ generations of
binary fussion occurred during this time. Every fifteen days, a sample of the 
population was frozen for later analysis giving rise to a `frozen fossil record'.
Since the population originated from a single cell, mutations, about $10^{6}$ everyday,
provided the sole source of genetic variation. Four years later, Lenski et al
had data for the evolving bacterial population over 10,000 generations.
They measured two quantities, the average cell size which is preserved in fossil
records, and the RF. They found that the average cell size
and the RF grow in a punctuated manner, i.e., in steps, as a function
of time (number of generations) when the data are plotted every 100 generations
(inset of Fig.1). At a larger interval of 500 generations, the changes 
appear to be gradual (inset of Fig.2 ). A major debate in evolutionary
biology revolves around the question of whether evolution is best described as
a gradual change or occurs in bursts. In the latter case, short periods of 
evolutionary activity are punctuated by long periods of stasis. This is the
theory of punctuated equilibrium (PE). Lenski et al's data seem to support
this theory though their interpretation is open to controversy \cite{Coyne}. The
experiment, however, clearly demonstrates that  both the average cell
size and the RF of the evolving population grow over a certain time interval. In the first 2000
generations or so, there is a rapid growth followed by a period of slower
growth till the growth is imperceptible. The bacteria, being in low sugar 
solution, have to compete for the food. Natural selection favours the 
mutations that confer some competitive advantage in exploiting the experimental
environment. This leads to adaptation of the bacterial population to the 
environment through the emergence of larger and fitter varients.

In the second experiment \cite{Lenski}, twelve populations of E.coli were evolved over 10,000
generations in identical environments. Each population was founded by a single
cell from an asexual clone to eliminate genetic variation within and between
populations. It was found that the replicate populations, after 10,000 generations,
differ considerably from one another in both the average cell size and the RF 
(inset of Fig.3 ), even though the populations evolved in identical
environments. In the third experiment \cite{Travisano}, the relative contributions of adaptation,
chance (mutations) and history to evolution were investigated. Twelve replicate
populations were founded from a single clone of E.coli and serially propagated
for 2000 generations in glucose-limited medium. The 12 populations had similar
fitness values when evolving in glucose medium but when put in a maltose-limited
medium showed large differences in fitness values. Some populations thrived 
while some others were found to languish. Ancestral fitness values of the 
populations in maltose were thus very heterogeneous (inset of Fig.4). One 
genotype from each of the 12 replicate populations was cloned and from this 
3 new replicate populations were founded. The 36 populations were then evolved 
under ancestral conditions in the maltose medium. The inset in Fig.4 shows
the derived fitness in maltose the versus ancestrial fitness in maltose for 36
populations.

Bak and Sneppen (BS) \cite{Bak,B} have proposed a model to study the evolution of
interacting species. The model self-organises into a critical state characterised by
power-laws of various types.
Further, evolutionary activity in this state exhibits PE. Boettcher and Paczuski
\cite{Bo} have proposed the M-trait BS model in which each biological species
is characterised by M traits, instead of one trait (fitness) as in the original
BS model. The major focus of these studies has been the self-organised critical
(SOC) state and its various features. The temporal evolution leading to the 
SOC state has not been studied in detail. In this letter, we show that, with 
some simple modifications of the M=2 BS model, many of the experimental results
of Lenski et al can be reproduced in a qualitative manner. Our study highlights
the significance of the dynamics in the BS-type models before the SOC state is
reached. The BS model gives a coarse-grained representation of real evolution
but contains the essential elements to capture the course of evolution. In our
modified M=2 BS model, the bacterial population is divided into N categories.
Each category contains bacteria of similar characteristics. The N categories 
correspond to the N sites of a one-dimensional (1D) lattice with periodic 
boundary conditions. In the original BS model, each site represents a species.
Two traits, namely, cell size and fitness are associated with the population at
each site i, i=1,2,...,N. One assigns a number (between 0 and 1) to each of the
traits at all the N sites. At each time step, the two sites with the minimum 
values for each of the two traits are identified. Mutations occur to bring about
changes in the traits. The minimum random numbers are replaced by new random
numbers. In the original M=2 BS model, the minimum value, amongst all the 2N 
values of the two traits, is replaced by a new random number. The random numbers
associated with any one of the traits of the neighbouring sites are also replaced
by new random numbers. This is to take into account the linkage of neighbouring
populations in food chain. The last two steps are repeated and averages are taken
for both the traits locally (over 40 sites) and globally (over 2000 sites). 
Unlike in the original BS model, we calculate quantities from the very beginning 
and not after the SOC state is reached. The SOC state corresponds to the region
in which evolutionary growth is imperceptible and fluctuates around an 
average value.
Fig.1  shows the variation of the
RF  versus time. The inset shows the experimental data 
\cite{Elena,Lenski}.
The averages are taken over 40 sites and every 100 time steps. The local averaging
gives rise to an improved quality of data points. Fig. 2
shows the corresponding variation with averages taken every 500 time steps
and over the whole lattice.  For a very large lattice one needs to take only global averages.
The RF is defined to be the ratio of the current 
fitness and the initial fitness at time t=0. In the actual experiment, fitness is
related to the growth rate of the bacterial population via replication.  
The RF  increases rapidly during the first 2000 
generations. After that the growth becomes slower till it becomes imperceptible. 
The rapid growths can be fit by an hyperbolic model
\begin{eqnarray}
y=x_{0}+\frac{ax}{b+x}
\end{eqnarray}
for both the experimental and simulation data. During the periods of punctuation
the beneficial mutations are few with no significant effect. When such mutations
occur in quick succession, rapid evolutionary growth is observed. Recent research findings \cite{VM} have highlighted
the importance of large beneficial mutations in the initial stages of evolutionary
growth. Organisms must adapt to the new conditions fairly quickly in order to
survive. Later, mutations with smaller effect fine-tune the adaptation. Fig.2
shows this clearly with a rapid evolutionary growth in the first 2000 generations
brought about by beneficial mutations of large effect. The growths are 
imperceptible when the bacterial population gets adapted to its environment.
The average cell size as a function of time has similar variations as in the case
of the RF (Figs. 1 and 2).

Fig.3 shows a comparison of the simulation data for the RF with the data (inset) 
of the second experiment of Lenski et al \cite{Lenski}. The plots show that the 
independent populations diverge significantly from one another . In the simulation, the initial random number
seed was chosen to be different for the six populations. The average fitness
 at time $t=0$ does not vary noticably from one
population to another. Fig.4 shows the simulation results for the third 
experiment \cite{Travisano} with $2\times4$ populations rather than the $3\times12$ populations
in the actual experiment. In the experiment, the populations growing in glucose-
limited medium were transferred after 2000 generations to maltose-limited medium.
In the latter medium, the average fitness values of the populations showed
large differences. Thus, in the simulation for the maltose medium, one starts
with widely different average fitness values for the populations. The populations
are evolved for 1000 time steps. One finds, in agreement with the experimental
results, that after 1000 generations, the average fitness values have similar
magnitudes. This shows that adaptation and chance (effect of mutation) have
eliminated the initial heterogeneity in average fitness values to a great
extent. The effect of history (initial heterogeneity) is reduced after several 
generations of evolution. The effect of adaptation is pronounced (shown by the
evolution of the data points above the isocline). The effect of chance is seen
in the small dispersion in the average fitness values of the two populations
corresponding to each genotype. Lenski et al \cite{Travisano} have given quantitative 
estimates of the relative contributions of adaptation, chance and history to 
average fitness and cell size before and after 1000 generations in maltose.

To conclude, we have described a set of experiments by Lenski et al on bacterial
evolution and have shown that a M=2 BS-type model gives a satisfactory description of the
experimental data. Both simulation and experiments support the theory of PE in
bacterial evolution. Recent exhaustive studies of fossil beds lend support to
this theory \cite{Kerr}.
The model is applied to evolving
bacterial populations rather than to interacting species as in the original BS 
model. Also, as emphasized earlier, the BS-type models have so far been studied
in the context of self-organised criticality. In this Brief Report, we highlight the
significance of the temporal evolution of BS-type models before the self-organised
critical state is reached. This temporal evolution needs to be characterised
in greater detail to give a quantitative fit to the experimental data of Lenski
et al. For example, appropriate change in random number intervals may be necessary
for a good fit. Lenski et al \cite{L,Gerrish,Joh,Vasi} have developed theories based on standard
population-genetics approaches to explain some of their experimental data.
The BS-type model gives a coarse-grained description to bacterial evolution
but, as shown in this Letter, it captures the major features of the evolutionary
growth of bacterial populations. Experiments have recently been performed on the
growth of RNA virus fitness \cite{No}. The adaptive evolutionary capacity in
this case is overwhelming. The gain in fitness is nearly 5000
after 50 passages. In that time, the gain in E.coli fitness changes can be 
explained by an hyperbolic model whereas RNA virus evolution 
follows exponential kinetics.
The RNA virus evolution 
follows exponential kinetics whereas E.coli evolution corresponds to hyperbolic
growth. A BS-type model can explain the hyperbolic growth but not the 
exponential one. For bacterial populations, Lenski et al's remarkable 
experiments open up possibilities of a rich interplay between theory and 
experiments. The self-organised critical state in the experiment and simulation
corresponds to the region in which growth is imperceptible. With appropriate designing
of experiments, the phenomenon of self-organised criticality in an evolving
bacterial population can be studied experimentally.

\section*{Acknowledgement}
We are greatful to R.E.Lenski for sending us his publications on bacterial
evolution.We also acknowledge the help from the Distributed Informatics
Centre, Bose Institute, in preparing the manuscript.

\newpage
\section*{Figure Captions}
\begin{description}
\item[Fig.1] Relative fitness versus time in experiment \cite{Lenski} and simulation.
A local average is taken over 40 sites in simulation. The experimental data
points are taken every 100 generations.
\item[Fig.2] Relative fitness versus time in simulation and experiment (inset)
\cite{Lenski}. The data points are taken every 500 generations and the average
is over the 2000 sites of the lattice.
\item[Fig.3] Trajectories for relative fitness in six replicate populations
of bacteria during 10,000 generations (simulation) and the same for twelve
replicate populations (experiment). The data points are taken at an interval
of 500 generations.
\item[Fig.4] Evolution of fitness during 1000 generations in maltose. Derived
versus ancestral values for relative fitness in 8 populations (simulation)
and the same for 36 populations (experiment). The different symbols indicate
the different progenitor genotypes.
\end{description}

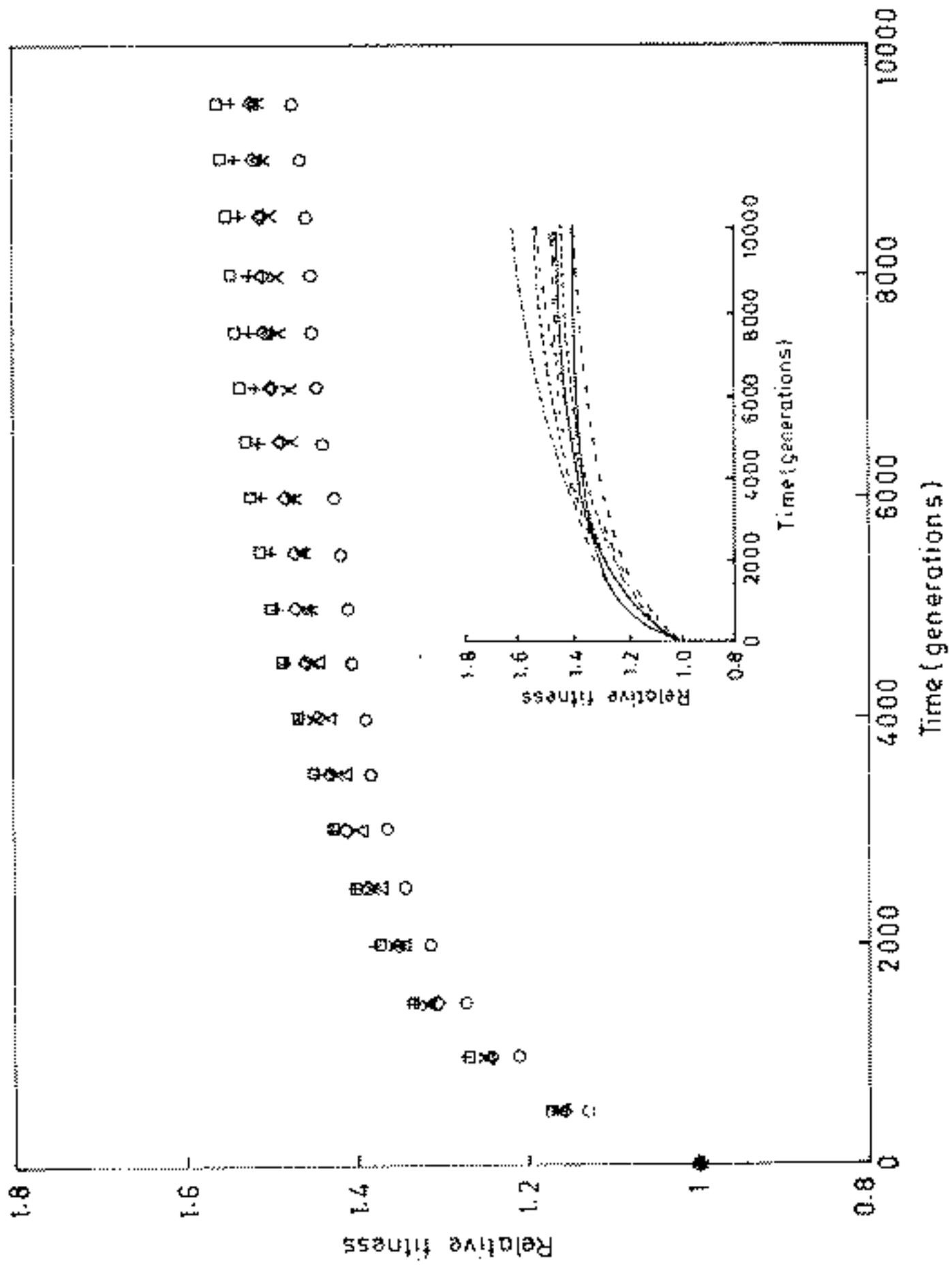
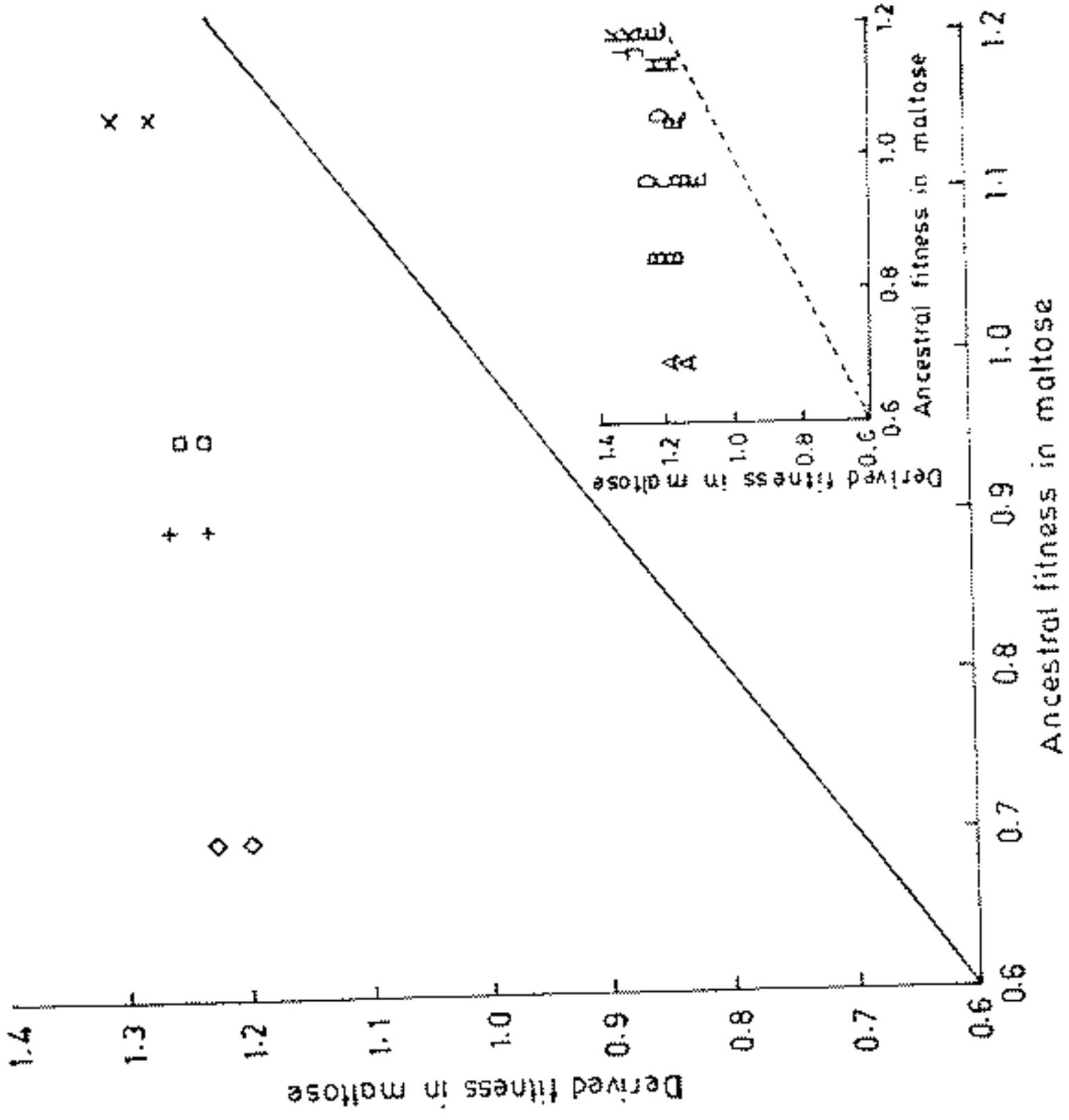
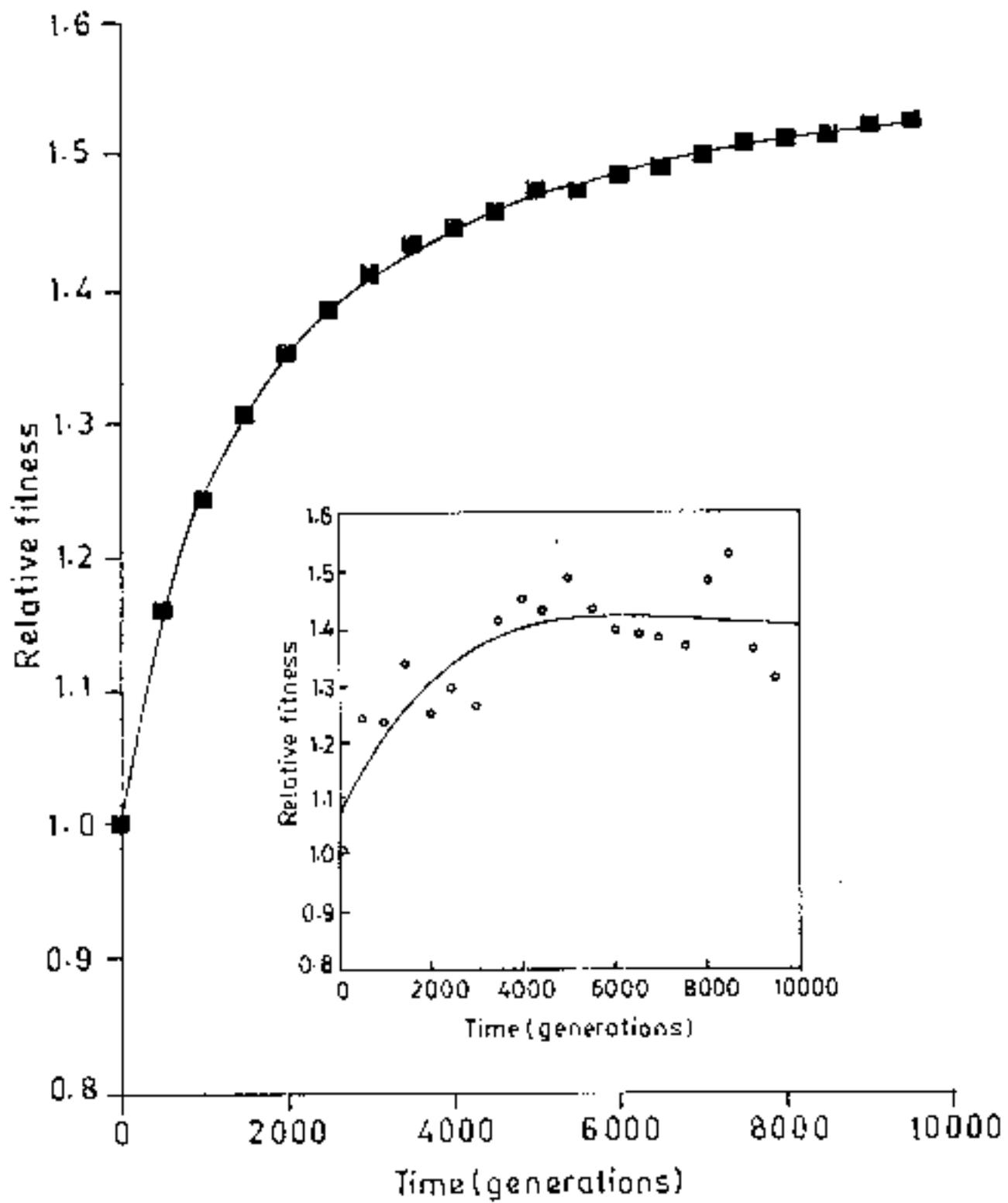
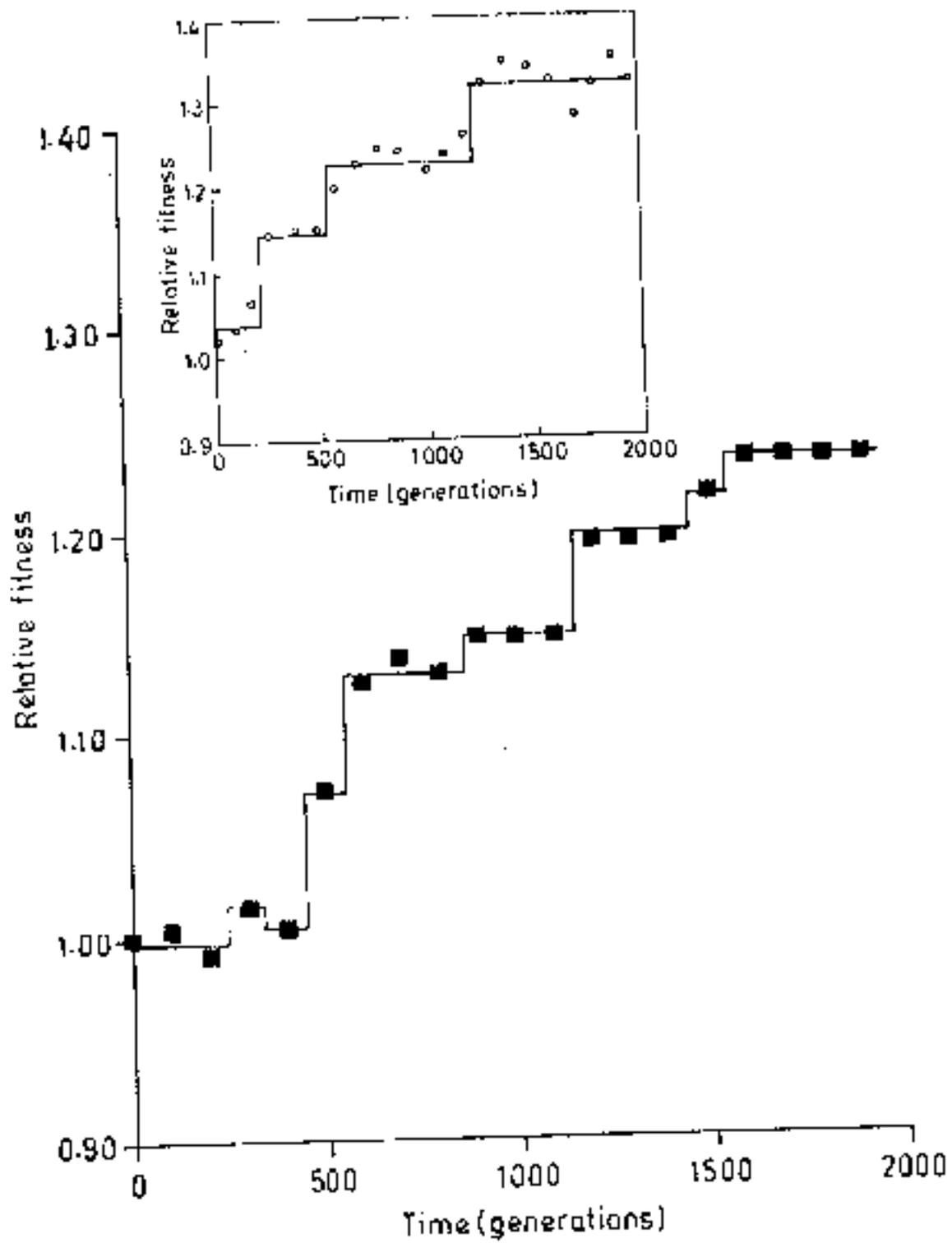
\end{document}